\documentclass[letterpaper,preprintnumbers,prd,nofootinbib,nobibnotes,showpacs]{revtex4}
\usepackage{amsfonts}
\usepackage{mathrsfs}
\usepackage{epsfig}
\usepackage{graphicx}%
\usepackage{dcolumn}
\usepackage{amsmath}
\usepackage{color}
\usepackage{color}
\makeatletter
\def\btt#1{\texttt{\@backslashchar#1}}%
\DeclareRobustCommand\bblash{\btt{\@backslashchar}}%
\makeatother
\begin{document}

\title{When the regularized Lovelock tensors are kinetically coupled to scalar field  }
\author{Changjun Gao}\email{gaocj@bao.ac.cn} \author{Shuang Yu}\email{yushuang@nao.cas.cn} \author{Jianhui Qiu}\email{jhqiu@nao.cas.cn}\affiliation{ Key Laboratory of Computational Astrophysics, National Astronomical Observatories, Chinese
Academy of Sciences, Beijing 100101, China}
\affiliation{University of Chinese Academy of Sciences, Beijing 100049, China}

\date{\today}

%%%%%%%%%%%%%%%%%%%%%%%%%%%%%%%%%%%%%%%%%%%%%%%%%%%%%%%%%%%%%%%%%%%%%%%%%%
\begin{abstract}
The recently proposed regularized Lovelock tensors are kinetically coupled to the scalar field. The resulting equation of motion is second order. In particular, it is found that when the $p=3$ regularized  Lovelock tensor is kinetically coupled to the scalar field, the scalar field is the potential candidate of cosmic dark energy.
\end{abstract}

% insert suggested PACS numbers in braces on next line
\pacs{04.70.Dy, 04.50.Gh, 05.70.Ce}

% insert suggested keywords - APS authors don't need to do this
%\keywords{}

\maketitle

%%%%%%%%%%%%%%%%%%%%%%%%%%%%%%%%%%%%%%%%%%%%%%%%%%%%%%%%%%%%%%%%%%%%%%%%%%
\section{Introduction}

Scientific breakthroughs are often manifested in making possible seeming impossible. As is well known, the Lovelock gravity is the most general gravity theory with second-order equations of motion \cite{ll:1971} constructed with the metric and its derivatives in $n$ dimensional spacetime . When $n=4$, the Lovelock gravity reduces to Einstein gravity. In other words, the higher orders of Lovelock tensors with $p\geq 2$ can not contribute to the equation of motion.

However, recently, a great breakthrough alters this perspective  \cite{glavan:2020,CasalinoCRV}. (See also \cite{tom:2011,cog:2013} for earlier works). The key idea of the breakthrough amounts to regularize the coupling constants
\begin{eqnarray}
\alpha_p\rightarrow \tilde{\alpha}_p=\alpha_p\frac{\left(n-2p-1\right)!}{\left(n-1\right)!}\;.
\end{eqnarray}
Then the Lovelock tensors $G^{(p)}_{\mu\nu}$ are regularized as
\begin{eqnarray}
G^{(p)}_{\mu\nu}\rightarrow \tilde{G}^{(p)}_{\mu\nu}=G^{(p)}_{\mu\nu}\frac{\left(n-2p-1\right)!}{\left(n-1\right)!}\;.
\end{eqnarray}
 It follows that the corresponding regularized Lovelock tensors $\tilde{G}^{(p)}_{\mu\nu}$ are non-vanishing, for example, in static spherically symmetric spacetime and maximally symmetric spacetime even if the spacetime is four dimensional.

The idea of this kind regularization was originally considered by Tomozawa \cite{tom:2011} with finite one-loop quantum corrections
to Einstein gravity.  Then it was also proposed by Cognola et al. \cite{cog:2013} with a classical Lagrangian approach.
With the invention of regularized Gauss-Bonnet tensor (or regularized $p=2$ Lovelock tensor) proposed by Glavan and Lin \cite{glavan:2020}, many interesting works have been carried out to investigate and extend the four dimensional Einstein-Gauss-Bonnet gravity.  These works include the exact solutions \cite{KumarGhoshMaharaj,SinghGM,DonevaYazadjiev,JusufiBG,GeSin,Liu14267,Yang14468,MaLu}, the black hole quasinormal modes \cite{KonoplyaZinhailoZhidenko,Churilova,Mishra,LiWY,ZhangZLG,AragonBGV,MalafarinaTD,Cuyubamba09025,LiuNZ,Devi14935}, the black hole shadows \cite{GuoLi,WeiLiu07769,ZhangWeiLiu,Heydari-Fard,RayimbaevATA,ZengZZ}, the gravitational lensing \cite{LiuZW,KumaraRHAA,IslamKG,JinGL,Kumar12970}, the black hole thermodynamics \cite{HegdeKA,SinghS,ZhangLG,Mansoori,WeiL14275,Konoplya02248,PanahJafarzade,YangWCYW,Ying}, the regularized Einstein-Gauss-Bonnet gravity \cite{LuPangMao,Kobayashi,Fernandes08362,HennigarKMP,BonifacioHJ}, the holographic superconductors \cite{qiao} and the lower-dimensional Gauss-Bonnet gravity \cite{robie}. Finally, some comments and objections can be found in \cite{Ai2020,GursesST,Mahapatra2020,Shu09339,TianZhu,ArrecheaDJ}.

The regularized Lovelock tensors $\tilde{G}^{(p)}_{\mu\nu}$ are derived by variations of the Lovelock action with respect to metric tensor. Therefore, they exactly obey the law of $\tilde{G}^{(p);\nu}_{\mu\nu}=0$. It is this nice property that enables the equation of motion for the scalar field is second order when the regularized Lovelock tensors are kinetically coupled. The corresponding Lagrangian is
\begin{eqnarray}
L&=&-\frac{1}{2}\sum_{p=0}^{N}\alpha_p \tilde{G}^{(p)}_{\mu\nu}\nabla^{\mu}\phi\nabla^{\nu}\phi-V\left(\phi\right)\;,
\end{eqnarray}
where $\alpha_p$ are constants and $V(\phi)$ is the scalar potential. The aim of this paper is to propose this coupling and investigate the cosmic evolution. We find that when the $p=3$ Lovelock tensor is kinetically coupled to the scalar field, the field can be the potential candidate of cosmic dark energy.

The paper is organized as follows. In Sec. II, we review the concept of regularized Lovelock tensors and propose its coupling with scalar field. In Sec. III, we study the cosmic evolution of the scalar field in the background of four dimensional and spatially flat Friedmann-Robertson-Walker Universe.  Then Sec. IV gives the conclusion and discussion. Throughout this paper, we adopt the system of units in which $G=c=\hbar=1$ and the metric signature
$(-,\ +,\ +,\ +)$.

\section{The regularized Lovelock tensors and its coupling with scalar field}\label{sec:2}
The action of Lovelock gravity \cite{ll:1971} takes the form
\begin{eqnarray}
{S}=\int d^n x\sqrt{-g}\left[\sum_{p}\alpha_p \lambda^{2\left(p-1\right)}L_{p}+L_m\right]\;,\label{ll}
\end{eqnarray}
where $n$ is the dimension of spacetime, $\alpha_p$ are constants and
summation is carried over all $p\in N$. $\lambda$ is a length scale. $L_m$ is the Lagrangian of matters.

$L_{p}$ is defined by
\begin{eqnarray}
L_{p}=2^{-p}\delta_{\sigma_1\sigma_2\cdot\cdot\cdot \sigma_{2p}}^{\lambda_1\lambda_2\cdot\cdot\cdot \lambda_{2p}}
R_{\lambda_1\lambda_2}^{\sigma_1\sigma_2}R_{\lambda_3\lambda_4}^{\sigma_3\sigma_4}\cdot\cdot\cdot R_{\lambda_{2p-1}\lambda_{2p}}^{\sigma_{2p-1}\sigma_{2p}}
\;,
\end{eqnarray}
where $\delta_{\sigma_1\sigma_2\cdot\cdot\cdot \sigma_{2p}}^{\lambda_1\lambda_2\cdot\cdot\cdot \lambda_{2p}}$
is the generalized Kronecker delta of the order $2p$. It equals to $\pm 1$ if the upper indices form
an even or odd permutation of the lower ones, respectively,
and zero in all other cases. Here $R_{\lambda_i\lambda_j}^{\sigma_k\sigma_l}$ is the Riemann tensor.

For example, we have
\begin{eqnarray}
L_{0}=1\;,\ \ \  L_{1}=R\;,\ \ \ \ L_{2}=R_{\mu\nu\alpha\beta}R^{\mu\nu\alpha\beta}-4R_{\mu\nu}R^{\mu\nu}+R^2\;.
\end{eqnarray}
They correspond the cosmological constant, Einstein-Hilbert Lagrangian and Lanczos Lagrangian \cite{lan:1932,lan:1938}, respectively.

The variation of action with respect to the metric gives the Lovelock gravity \cite{ll:1971}
\begin{eqnarray}
\sum_{p}\alpha_p\lambda^{2\left(p-1\right)}G_{(p)\mu\nu}=\kappa T_{\mu\nu}\;.
\end{eqnarray}
$T_{\mu\nu}$ is the energy momentum tensor of matters and $\kappa$ is a constant. In four dimensional case, $\kappa=8\pi$. The Lovelock tensors are
\begin{eqnarray}
G^{\mu}_{(p)\nu}=-2^{-p-1}\delta_{\nu\sigma_1\sigma_2\cdot\cdot\cdot \sigma_{2p}}^{\mu\lambda_1\lambda_2\cdot\cdot\cdot \lambda_{2p}}
R_{\lambda_1\lambda_2}^{\sigma_1\sigma_2}R_{\lambda_3\lambda_4}^{\sigma_3\sigma_4}\cdot\cdot\cdot R_{\lambda_{2p-1}\lambda_{2p}}^{\sigma_{2p-1}\sigma_{2p}}\;.
\end{eqnarray}

In particular, we have \cite{bri:1997}
\begin{eqnarray}
&&G^{(0)}_{\mu\nu}=-\frac{1}{2}g_{\mu\nu}\;,\nonumber\\
&&G^{(1)}_{\mu\nu}=R_{\mu\nu}-\frac{1}{2}g_{\mu\nu}R\;,\nonumber\\
&&G^{(2)}_{\mu\nu}=-\frac{1}{2}g_{\mu\nu}\left(R^2-4R_{\kappa\sigma}R^{\kappa\sigma}+R_{\kappa\sigma\tau\rho}
R^{\kappa\sigma\tau\rho}\right)+2\left(RR_{\mu\nu}-R_{\mu\sigma\kappa\tau}R^{\kappa\tau\sigma}_{\nu}-2R_{\mu\kappa\nu\sigma}R^{\kappa\sigma}
-2R_{\mu\sigma}R^{\sigma}_{\nu}\right)\;,\nonumber\\
&&G^{(3)}_{\mu\nu}=\frac{1}{2}g_{\mu\nu}\left(12RR_{\kappa\sigma}R^{\kappa\sigma}-R^3-3RR_{\alpha\beta\sigma\kappa}R^{\alpha\beta\sigma\kappa}
-16R_{\alpha}^{\beta}R_{\beta}^{\sigma}R_{\sigma}^{\alpha}+24R_{\alpha\beta}R_{\sigma\kappa}R^{\alpha\sigma\beta\kappa}+24R_{\alpha}^{\beta}
R^{\alpha\sigma\kappa\rho}R_{\beta\sigma\kappa\rho}\right.\nonumber\\&&\left.+2R_{\alpha\beta}^{\sigma\kappa}R_{\sigma\kappa}^{\rho\lambda}R_{\rho\lambda}^{\alpha\beta}
-8R_{\alpha\beta}^{\sigma\kappa}R_{\sigma\rho}^{\alpha\lambda}R_{\kappa\lambda}^{\beta\rho}\right)-24R_{\mu\alpha\beta\sigma}
R_{\nu}^{\beta}R^{\alpha\sigma}
-12R_{\mu\nu}R_{\alpha\beta}R^{\alpha\beta}+24R_{\mu}^{\alpha}R_{\alpha}^{\beta}R_{\beta\nu}
+24R_{\mu}^{\alpha}R^{\beta\sigma}R_{\alpha\beta\sigma\nu}\nonumber\\&&+3R_{\mu\nu}R^2+3R_{\mu\nu}R_{\alpha\beta\sigma\kappa}R^{\alpha\beta\sigma\kappa}
-12R_{\mu\alpha}R_{\nu\beta\sigma\kappa}R^{\alpha\beta\sigma\kappa}
+6RR_{\mu\alpha\beta\sigma}R_{\nu}^{\alpha\beta\sigma}-24R_{\mu\alpha\nu\beta}R_{\sigma}^{\alpha}R^{\sigma\beta}-12RR_{\mu}^{\sigma}R_{\sigma\nu}\nonumber\\&&+24R_{\mu\alpha\nu\beta}R_{\sigma\kappa}R^{\alpha\sigma\beta\kappa}-12R_{\mu\alpha\beta\sigma}
R^{\kappa\alpha\beta\sigma}R_{\kappa\nu}-12R_{\mu\alpha\beta\sigma}R^{\alpha\kappa}R_{\nu\kappa}^{\beta\sigma}
+12RR_{\mu\sigma\nu\kappa}R^{\sigma\kappa}\nonumber\\&&+12R_{\mu\alpha\nu\beta}R^{\alpha}_{\sigma\kappa\rho}R^{\beta\sigma\kappa\rho}+
6R_{\mu}^{\alpha\beta\sigma}R_{\beta\sigma}^{\kappa\rho}R_{\kappa\rho\alpha\sigma}+24R_{\mu\alpha}^{\beta\sigma}R_{\beta\nu\rho\lambda}
R_{\sigma}^{\lambda\alpha\rho}+24R_{\mu}^{\alpha\beta\sigma}R_{\beta}^{\kappa}R_{\sigma\kappa\nu\beta}\;.
\end{eqnarray}
In the background of Friedmann-Robertson-Walker Universe
\begin{eqnarray}
ds^2&=&-dt^2+a\left(t\right)^2\left(dr^2+r^2d\Omega_{n-2}^2\right)\;,
\end{eqnarray}
where $a(t)$ is the scale factor of the Universe,
the Lovelock Lagrangians and tensors reduce to \cite{far:1990}

\begin{eqnarray}
L_p&=&H^{2p}\frac{n!}{\left(n-2p\right)!}\;,\ \ \ \ G_{(p)\mu\nu}=\frac{1}{2}H^{2p}\frac{\left(n-1\right)!}{\left(n-2p-1\right)!}g_{\mu\nu}\;,
\end{eqnarray}
with
\begin{eqnarray}
H=\dot{a}/a\;,
\end{eqnarray}
the Hubble parameter. It is apparent we  must have
\begin{eqnarray}
0\leq 2p<n-1\;.
\end{eqnarray}
Then we conclude that when $n=4$, we have $p=0$ and $p=1$. In other words, only the cosmological term and Einstein tensor are non-vanishing. In order that all the Lovelock tensors make contributions in the equations of motion for arbitrary dimensions, Caslina et al.  propose the regularized Lovelock gravity \cite{CasalinoCRV} by substitution
\begin{eqnarray}
\alpha_p\rightarrow \alpha_p\frac{\left(n-2p-1\right)!}{\left(n-1\right)!}\;.
\end{eqnarray}
By this substitution, the Lovelock Lagrangians and tensors become
\begin{eqnarray}
\tilde{L}_p&=&H^{2p}\frac{n}{\left(n-2p\right)}\;,\ \ \ \ \tilde{G}_{(p)\mu\nu}=\frac{1}{2}H^{2p}g_{\mu\nu}\;.
\end{eqnarray}
Now all the orders of Lagrangians and tensors are non-vanishing in four dimensional spacetime because they are regularized.
Since the regularized Lovelock tensors $\tilde{\tilde{G}}^{(p)}_{\mu\nu}$ are derived by variations of the action, they exactly obey the law of $\tilde{G}^{(p);\nu}_{\mu\nu}=0$. It is this desired property that enables us to kinetically couple it with scalar field
\begin{eqnarray}
L&=&-\frac{1}{2}\sum_{p=0}^{N}\alpha_p \tilde{G}^{(p)}_{\mu\nu}\nabla^{\mu}\phi\nabla^{\nu}\phi-V\left(\phi\right)\;,
\end{eqnarray}
where $\alpha_p$ are constants and $V(\phi)$ is the scalar potential.

The equation of motion for the scalar field is
\begin{equation}
\sum_{p}^{N}\alpha_p \tilde{G}^{(p)}_{\mu\nu}\nabla^{\mu}\nabla^{\nu}\phi-\frac{dV}{d\phi}=0\;.
\end{equation}
It is an equation of motion with second order derivative. For simplicity, we shall consider $V=0$ in the next section.

\section{cosmic evolution}
\label{sec:3}
In this section, we investigate the cosmic evolution of the scalar field in the $4$ dimensional spatially flat Friedmann-Robertson-Walker Universe. The metric is given by
 \begin{eqnarray}
ds^2&=&-dt^2+a\left(t\right)^2\left(dr^2+r^2d\Omega_2^2\right)\;,
\end{eqnarray}
where $a(t)$ is the scale factor of the Universe. Taking into account the contribution of ordinary matter, dark matter and radiation matter, we have the total action as follows
\begin{eqnarray}
{S}=\int d^4 x\sqrt{-g}\left[\frac{R}{16\pi}-\frac{1}{2}\sum_{p}\alpha_p \tilde{G}^{(p)}_{\mu\nu}\nabla^{\mu}\phi\nabla^{\nu}\phi+L_{m}\right]\;.
\end{eqnarray}
After tedious calculations, we obtain the Friedmann equation
\begin{eqnarray}\label{eom1}
3H^2=8\pi\left[\frac{1}{2}\sum_{p=0}^{N}\alpha_p \left(p+\frac{1}{2}\right)H^{2p}\dot{\phi}^2+\rho_m\right]\;,
\end{eqnarray}
the acceleration equation
\begin{eqnarray}\label{eom1}
2\dot{H}+3H^2=-8\pi\left\{\frac{1}{2}\sum_{p=0}^{N}\left[\frac{1}{3}\alpha_p pH^{2p+1}\dot{\phi}\ddot{\phi}+\frac{1}{12}\alpha_p H^{2p}\left(4p^2\dot{H}-2p\dot{H}+6pH^2-3H^2\right)\dot{\phi}^2\right]+p_m\right\}\;,
\end{eqnarray}
and the equation of motion for scalar field
\begin{eqnarray}\label{eoms}
\left(\sum_{p=0}^{N}\alpha_p a^3 H^{2p}\dot{\phi}\right)^{\cdot}=0\;,
\end{eqnarray}
where dot denotes the derivative with respect to cosmic time $t$. $\rho_m$ and $p_m$ are the energy densities and pressures of matters which mainly cover dark matter and radiation matter. We note that only two of the above three equations are independent.

By solving Eq.~(\ref{eoms}), we find the Friedmann equation can be written as
\begin{eqnarray}\label{Fried}
3H^2=8\pi\left[\frac{1}{a^6F^2}\frac{d\left(FH\right)}{dH}+\rho_m\right]\;,
\end{eqnarray}
 where $F$ is defined by
\begin{eqnarray}\label{eom1}
F\equiv \sum_{p}^{N}\alpha_p H^{2p} \;.
\end{eqnarray}
From the Friedmann equation (\ref{Fried}), the energy density of scalar field is read out
 \begin{eqnarray}\label{eom1}
\rho_X=\frac{1}{a^6F^2}\frac{d\left(FH\right)}{dH}\;.
\end{eqnarray}
It is the function of Hubble parameter. In particular, when
\begin{eqnarray}
F=\alpha_p H^{2p}\;,
\end{eqnarray}
which comes from the coupling of $p$-th order Lovelock tensor with scalar field, we have
\begin{eqnarray}\label{eom1}
\rho_X=\frac{2p+1}{\alpha_p a^6H^{2p}}\;.
\end{eqnarray}
It tells us that, in the radiation dominated
epoch of the Universe, we have $H^2\propto a^{-4}$ such that $\rho_X\propto a^{4p-6}$. For $p\geq 2$, the scalar field behaves as a phantom dark energy, growing with the expansion of Universe. On the other hand, in the
matter dominated epoch, we have $H^2\propto a^{-3}$ such that $\rho_X\propto a^{3p-6}$. For the $p=2$ Lovelock tensor, $\rho_X\approx cost$. In this case, the scalar field behaves as a cosmological constant. But the detailed analysis in the next shows that $p=3$ Lovelock tensor coupling is the potential candidate of cosmic dark energy.

By convention, the Friedmann equation can be reformulated
as
\begin{eqnarray}
1={\Omega_X}+{\Omega_d}+{\Omega_{r}}\;,
\end{eqnarray}
with
\begin{eqnarray}
\Omega_X\equiv\frac{\Omega_{0X}}{a^6 h^{2p}}\;, \ \ \ \ \Omega_d\equiv\frac{\Omega_{0d}}{a^3h^2}\;,\ \ \  \Omega_{r}\equiv\frac{\Omega_{0r}}{a^4 h^2}\;.
\end{eqnarray}
Here $h={H}/{H_0}$ and $H_0$ are the normalized and present-day Hubble parameters, respectively. $\Omega_d$, $\Omega_r$ and $\Omega_X$ are the ratios of various components. Astrophysical observations suggest the present-day ratios are $\Omega_{0d}=0.25$, $\Omega_{0r}=10^{-4}$ and $\Omega_{0X}=0.75$.

\begin{figure}[htbp]
	\centering
	\includegraphics[width=8cm,height=6cm]{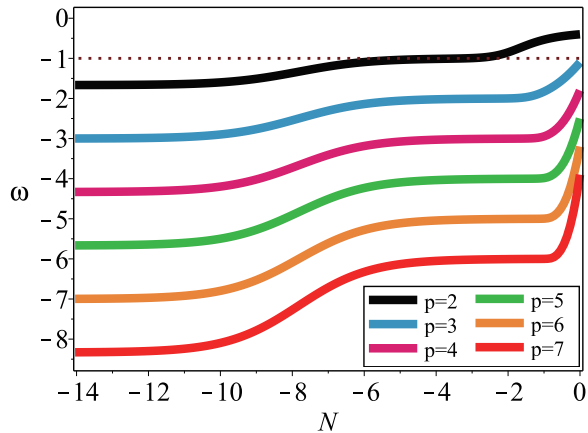}
	\caption{The cosmic evolution of equation of state for the scalar field. They correspond to $p=2,\ 3,\  4,\ 5,\ 6$ from top to bottom. The case of $p=3$ is preferable to being the candidate of cosmic dark energy.}\label{eos-2}
\end{figure}

\begin{figure}[htbp]
	\centering
	\includegraphics[width=8cm,height=6cm]{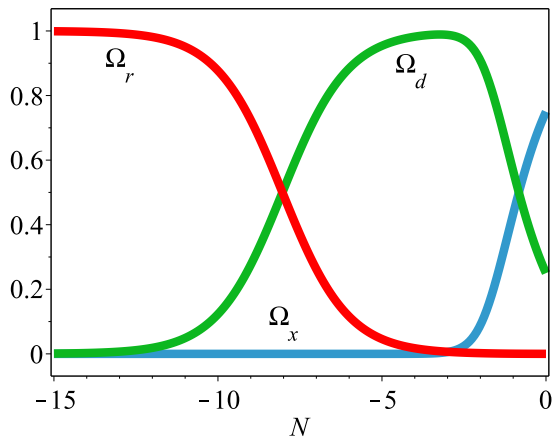}
	\caption{The cosmic evolution of ratios for dark matter $\Omega_{d}$, radiation matter $\Omega_{r}$ and the scalar field $\Omega_X$ for the coupling of $p=3$ regularized Lovelock tensor.}\label{density1}
\end{figure}

In Fig.~\ref{eos-2}, we plot the equation of state $\omega$
\begin{eqnarray}
 \omega=-\frac{1}{3}\frac{d\ln{\rho_X}}{d\ln{a}}-1\;,
\end{eqnarray}
with respect to $N=\ln{a}$ for $p=2,\ 3,\  4,\ 5,\ 6$ from top to bottom. Observations reveal that the present-day equation of state for dark energy is around $-1$. For $p=3$, we have the present-day equation of state $\omega_0=-1.1$. Thus the case of $p=3$ is preferable to being the candidate of dark energy.
In Fig.~\ref{density1}, we plot the cosmic evolution of ratios for dark matter, radiation matter and the scalar field with respect to $N=\ln{a}$. They are consistent with the observations very well.

\section{conclusion and discussion}
It is well known that the Lovelock tensors with $p\geq 2$ are vanishing when the spacetime is four dimensional. This property seriously blocks their application in four dimensional spacetime. However, if the Lovelock gravity is regularized \cite{CasalinoCRV}, all the orders of Lovelock tensors are non-vanishing irrespective of the dimension of spacetime. Thus this regularization clears the way for the research of Lovelock tensors in physical spacetime.
In essence, the regularized Lovelock tensors are derived by the variation of gravitational action with respect to metric tensor. Therefore, they exactly obey the law of $\tilde{G}^{(p);\nu}_{\mu\nu}=0$. Namely, the covariant divergence is vanishing. It is this nice property that enables the equation of motion for the scalar field is second order when the regularized tensors are kinetically coupled.

Then the cosmic evolution of the scalar field is investigated. The Friedmann equation and the acceleration equation are derived. The energy density of the scalar
field is the positive series of Hubble parameter
\begin{eqnarray}\label{eom1}
F\equiv \sum_{p=0}^{N}\alpha_p H^{2p} \;.
\end{eqnarray}
In principle, one can let $N=+\infty$ since the order $p$ of the Lovelock tensor can be arbitrarily large. Then the energy density becomes
\begin{eqnarray}\label{eom1}
F\equiv \sum_{p=0}^{+\infty}\alpha_p H^{2p} \;.
\end{eqnarray}
It is an infinite series of Hubble parameter. So we can get any form of $F(H)$ we want if the coupling constants $\alpha_p$ are properly chosen. For example, $F=e^{-\alpha H^6}$ embraces the contribution of all the Lovelock tensors with $p\geq 3$. We have studied the equation of state for the scalar field when $p=2,\ 3,\,4,\ 5,\ 6$. It is shown that when $p\geq3$, the scalar field behaves as phantom dark energy, growing with the expansion of the Universe. In particular, the case of $p=3$ can be the potential candidate of cosmic dark energy.

Anyway, the invention of regularized Lovelock tensors opens a new window for the research in physics. Many interesting phenomena need to be discovered. For example, when they are kinetically coupled to scalar field,
the case of $p=3$ is the potential candidate of cosmic dark energy.

\section*{Acknowledgments}
This work is partially supported by China Program of International ST Cooperation 2016YFE0100300
, the Strategic Priority Research Program ``Multi-wavelength Gravitational Wave Universe'' of the
CAS, Grant No. XDB23040100, the Joint Research Fund in Astronomy (U1631118), and the NSFC
under grants 11473044, 11633004, 11773031 and the Project of CAS, QYZDJ-SSW-SLH017.

\newcommand\ARNPS[3]{~Ann. Rev. Nucl. Part. Sci.{\bf ~#1}, #2~ (#3)}
\newcommand\AL[3]{~Astron. Lett.{\bf ~#1}, #2~ (#3)}
\newcommand\AP[3]{~Astropart. Phys.{\bf ~#1}, #2~ (#3)}
\newcommand\AJ[3]{~Astron. J.{\bf ~#1}, #2~(#3)}
\newcommand\APJ[3]{~Astrophys. J.{\bf ~#1}, #2~ (#3)}
\newcommand\APJL[3]{~Astrophys. J. Lett. {\bf ~#1}, L#2~(#3)}
\newcommand\APJS[3]{~Astrophys. J. Suppl. Ser.{\bf ~#1}, #2~(#3)}
\newcommand\JHEP[3]{~JHEP.{\bf ~#1}, #2~(#3)}
\newcommand\JMP[3]{~J. Math. Phys. {\bf ~#1}, #2~(#3)}
\newcommand\JCAP[3]{~JCAP {\bf ~#1}, #2~ (#3)}
\newcommand\LRR[3]{~Living Rev. Relativity. {\bf ~#1}, #2~ (#3)}
\newcommand\MNRAS[3]{~Mon. Not. R. Astron. Soc.{\bf ~#1}, #2~(#3)}
\newcommand\MNRASL[3]{~Mon. Not. R. Astron. Soc.{\bf ~#1}, L#2~(#3)}
\newcommand\NPB[3]{~Nucl. Phys. B{\bf ~#1}, #2~(#3)}
\newcommand\CMP[3]{~Comm. Math. Phys.{\bf ~#1}, #2~(#3)}
\newcommand\CQG[3]{~Class. Quantum Grav.{\bf ~#1}, #2~(#3)}
\newcommand\PLB[3]{~Phys. Lett. B{\bf ~#1}, #2~(#3)}
\newcommand\PRL[3]{~Phys. Rev. Lett.{\bf ~#1}, #2~(#3)}
\newcommand\PR[3]{~Phys. Rep.{\bf ~#1}, #2~(#3)}
\newcommand\PRd[3]{~Phys. Rev.{\bf ~#1}, #2~(#3)}
\newcommand\PRD[3]{~Phys. Rev. D{\bf ~#1}, #2~(#3)}
\newcommand\RMP[3]{~Rev. Mod. Phys.{\bf ~#1}, #2~(#3)}
\newcommand\SJNP[3]{~Sov. J. Nucl. Phys.{\bf ~#1}, #2~(#3)}
\newcommand\ZPC[3]{~Z. Phys. C{\bf ~#1}, #2~(#3)}
\newcommand\IJGMP[3]{~Int. J. Geom. Meth. Mod. Phys.{\bf ~#1}, #2~(#3)}
\newcommand\IJMPD[3]{~Int. J. Mod. Phys. D{\bf ~#1}, #2~(#3)}
\newcommand\GRG[3]{~Gen. Rel. Grav.{\bf ~#1}, #2~(#3)}
\newcommand\EPJC[3]{~Eur. Phys. J. C{\bf ~#1}, #2~(#3)}
\newcommand\PRSLA[3]{~Proc. Roy. Soc. Lond. A {\bf ~#1}, #2~(#3)}
\newcommand\AHEP[3]{~Adv. High Energy Phys.{\bf ~#1}, #2~(#3)}
\newcommand\Pramana[3]{~Pramana.{\bf ~#1}, #2~(#3)}
\newcommand\PTP[3]{~Prog. Theor. Phys{\bf ~#1}, #2~(#3)}
\newcommand\APPS[3]{~Acta Phys. Polon. Supp.{\bf ~#1}, #2~(#3)}
\newcommand\ANP[3]{~Annals Phys.{\bf ~#1}, #2~(#3)}

\end{document}